\documentclass[5p, twocolumn, 10pt, sort&compress]{elsarticle}

\usepackage{amssymb}
\usepackage{amsthm}

\usepackage{lineno}

\biboptions{comma,square}

\newcounter{bla}

\journal{Computer Physics Communications}

\usepackage{amsmath}
\usepackage{mathtools}
\usepackage{derivative}
\usepackage{graphicx, dcolumn, bm}
\usepackage[hidelinks]{hyperref}
\usepackage[dvipsnames]{xcolor}
\usepackage{orcidlink}
\usepackage{seqsplit}
\usepackage{environ}

\hypersetup{
colorlinks=true
}

\usepackage{newfloat}
\usepackage[frozencache, cachedir=./listingCache]{minted}
\setminted{
bgcolor = gray!20,
breaklines,
python3
}
\newcommand{\inlinecode}[1]{\mintinline[breaklines, bgcolor=gray!20, python3]{python}{#1}}

\newcommand{\matrixel}[3]{\left\langle{#1}\middle|{#2}\middle|{#3}\right\rangle}
\newcommand{\trace}[1]{\mathrm{tr}\left({#1}\right)}
\newcommand{\ket}[1]{\left|{#1}\right\rangle}

\newcommand{\expval}[1]{\left\langle{#1}\right\rangle}
\newcommand{\comm}[2]{\left[{#1},{#2}\right]}
\newcommand{\acomm}[2]{\left\{{#1},{#2}\right\}}

\newcommand{\dissip}[1]{\mathcal{D}\left({#1}\right)\left[\rho\right]}

\newcommand{\bop}{\hat{b}}

\newcommand{\bdag}{\bop^\dagger}
\newcommand{\bdagn}[1]{\bop^{\dagger {#1}}}

\newcommand{\pybolano}{\texttt{pyBoLaNO}~}

\newenvironment{revision}{}{}

\newenvironment{revision2}{%
\color{red}
}
{}

\begin{document}
\sloppy

\begin{frontmatter}

\title{\texttt{pyBoLaNO}: A \texttt{Python} symbolic package for normal ordering involving bosonic ladder operators}

\author[a,b]{Hendry M. Lim\corref{author}}
\author[c,a]{Donny Dwiputra}
\author[a]{M. Shoufie Ukhtary}
\author[a,d]{Ahmad R. T. Nugraha}

\cortext[author] {Corresponding author.\\\textit{E-mail address:} hendry01@ui.ac.id}
\address[a]{Research Center for Quantum Physics, National Research and Innovation Agency (BRIN), South Tangerang 15314, Indonesia}
\address[b]{Department of Physics, Faculty of Mathematics and Natural Sciences, Universitas Indonesia, Depok 16424, Indonesia}
\address[c]{Asia Pacific Center for Theoretical Physics, Pohang 37673, Korea}
\address[d]{Department of Engineering Physics, Telkom University, Bandung 40257, Indonesia}


\begin{abstract}
We present \texttt{pyBoLaNO}, a \texttt{Python} symbolic package based on \texttt{SymPy} to quickly normal-order any polynomial in bosonic ladder operators \begin{revision}regarding the canonical commutation relations, using Blasiak's formulae\end{revision}. By extension, this package offers the normal ordering of commutators of any two polynomials in bosonic ladder operators and the evaluation of the normal-ordered expectation value evolution in the Lindblad master equation framework for open quantum systems. The package supports multipartite descriptions and multiprocessing. We describe the package's workflow, show examples of use, and discuss its computational performance. All codes and examples are available on our GitHub repository.
\end{abstract}



\end{frontmatter}


\noindent {\bf PROGRAM SUMMARY}

\begin{small}
\noindent
{\em Program Title:} \texttt{pyBoLaNO}
\\
{\em Developer's repository link:} \href{https://github.com/hendry24/pyBoLaNO}{\seqsplit{https://github.com/hendry24/pyBoLaNO}}
\\
{\em Licensing provisions:} MIT License
\\
{\em Programming language:} \texttt{Python}
\\
{\em Nature of problem:} Normal ordering involving bosonic ladder operators \begin{revision} regarding the canonical commutation relations. \end{revision}
\\
{\em Solution method:} Blasiak's formulae for the normal ordering of an arbitrary monomial in bosonic ladder operators \begin{revision}regarding the canonical commutation relations.\end{revision} Symbolic programming is fully provided by \texttt{SymPy}.
\end{small}


\section{Introduction}
\label{Introduction}

Ladder operators arise in the study of the quantum simple harmonic oscillator. They consist of the annihilation/lowering operator $\bop$ and its Hermitian conjugate, the creation/raising operator $\bdag$. They allow for the algebraic treatment of the problem, giving a simple relation between the system's eigenstates. The eigenstate $\ket{n}$ corresponding to the $n$th energy level $E_n$ is obtained by applying $\hat{b}^\dagger$ to the vacuum state $\ket{0}$ a total of $n$ times, $\ket{n}\propto\hat{b}^{\dagger n}\ket{0}$~\cite{griffiths_introduction_2018, lancaster_quantum_2014}. Taking into account quantum statistics, the ladder operators are defined differently for bosons and fermions---usually denoted $\hat{a}$ and $\hat{a}^\dagger$ in the latter case. Furthermore, the eigenvectors of the bosonic annihilation operator are quantum states that most closely resemble the classical harmonic oscillator, that is, the coherent states~\cite{gerry_introductory_2005}. The formulation is far-reaching in quantum mechanics. The ladder operators form the basis for quantum field theory, in which a particle is considered as the excitation of the underlying quantum field~\cite{lancaster_quantum_2014}. They can also be found in the theoretical description of physical systems in atomic, molecular, and optical physics~\cite{gerry_introductory_2005} that extends beyond simple harmonic oscillators.

\begin{revision}We are interested in obtaining the normal-ordered equivalent of a polynomial in the ladder operators---a process we shall call ``normal ordering'' herein. This is useful in quantum optics as physically relevant expectation values involve normal-ordered monomials in the ladder operators. Matrix elements in the coherent-state basis can be straightforwardly evaluated for normal-ordered monomials. Furthermore, given the system's Glauber-Sudarshan $P$ function, the normal-ordered form is convenient via the optical equivalence theorem~\cite{gerry_introductory_2005, fox2006quantum}. It is also quite common in the literature to present the evolution equation for a given expectation value in a normal-ordered form~\cite{chia_relaxation_2020, Shen2023, downing_hyperbolic_2024, Krimer2019, Zens2019, Ahmadi2024, Downing2021, Downing2022, Downing2023, BenArosh2021, Amitai2018.PhysRevE.97.052203, Minganti2019, Chimczak2023, Farina2019, Zhang2021}.\end{revision} As the system description becomes more complex, the algebraic manipulation of the dynamical equations becomes more tedious and prone to errors, making automation desirable.

The \texttt{SymPy} package~\cite{meurer_sympy_2017} provides a symbolic computation framework in the \texttt{Python} programming language. \begin{revision}At the time of writing this work, \texttt{SymPy} is on release \inlinecode{ver. 1.13.3} which supports the ladder operators via the \inlinecode{sympy.physics.secondquant} and \inlinecode{sympy.physics.quantum} submodules. In particular, normal ordering is implemented as the function \inlinecode{normal_ordered_form}. \emph{Unfortunately}, the algorithm implemented in the package quickly slows down as we increase the complexity of the expression to normal-order.\end{revision}

\begin{revision}Motivated by a more efficient method and the need for a convenient tool to obtain the expectation value evolution given the equation of motion, we develop the \texttt{Python} package \pybolano that offers fast symbolic normal ordering of expressions involving bosonic ladder operators, which extends to fast normal ordering of commutators and expectation value evolution from the Lindblad master equation. The package is fully built on the core functionalities of \texttt{Sympy} and supports multipartite descriptions as well as multiprocessing for each additive term in the input(s).\end{revision}

The remainder of this paper is structured as follows. In Section~\ref{section:theoretical_considerations}, we elaborate on the theoretical basis for the features presented in this package. Section~\ref{section:package_anatomy} describes the package's functionalities. The package usage is shown through some selected examples in Section~\ref{section:examples}. The computational performance of the package is discussed in Section~\ref{section:performance}. Finally, Section~\ref{section:conclusion} concludes this paper.


\section{Theoretical Considerations}
\label{section:theoretical_considerations}

Here, we present a brief overview of the formulations of the ladder operators available in quantum mechanics textbooks. The readers interested in only the essentials of the package may skip to Section~\ref{subsec:normal_ordering}. 

We consider the classical Hamiltonian of the simple harmonic oscillator under Hooke's law~\cite{griffiths_introduction_2018}:
\begin{equation}
    H = \frac{p^2}{2m} + \frac{m\omega_0^2x^2}{2},
\end{equation}
where $m$ is the oscillator's mass and $\omega_0$ is its natural angular frequency. The quantization of this Hamiltonian is achieved by replacing the position $x$ and momentum $p$ with the corresponding Hilbert space operators. We obtain
\begin{equation}\label{eq:SHO_Ham}
    \hat{H} = \frac{\hat{p}^2}{2m} + \frac{m\omega_0^2\hat{x}^2}{2}.
\end{equation}
The time-independent Schr\"{o}dinger equation (TISE) is given by
\begin{equation}
    \hat{H}\psi = E\psi,
\end{equation}
where $\psi$ is the system's wave function and $E$ is its energy.  In the algebraic treatment of the problem (the other being the analytic method involving Hermite polynomials), we define the operators
\begin{subequations}
\begin{align}
    \bop &= \frac{m\omega_0 \hat{x}+i\hat{p}}{\sqrt{2\hbar m\omega_0}},
    \\
    \bdag &= \frac{m\omega_0 \hat{x}-i\hat{p}}{\sqrt{2\hbar m\omega_0}},
\end{align}
\end{subequations}
satisfying the commutation relations
\begin{align}
    \comm{\bop}{\bdag}&=1,
\end{align}
where $\comm{\hat{A}}{\hat{B}}=\hat{A}\hat{B}-\hat{B}\hat{A}$.  The Hamiltonian becomes
\begin{equation}
    \hat{H} = \hbar\omega_0\left(\bdag\bop+\frac{1}{2}\right).
\end{equation}
Considering the time-independent Schr\"{o}dinger equation (TISE),
\begin{equation}
    \hat{H}\ket{\psi} = E\ket{\psi},
\end{equation}
where $\psi$ is the wave function and $E$ is the energy, it can be shown that 
\begin{subequations}
    \begin{align}
        \hat{H}\left(\bdag\ket{\psi}\right) &= \left(E+\hbar\omega_0\right)\left(\bdag\ket{\psi}\right),
        \\
        \hat{H}\left(\bop\ket{\psi}\right) &= \left(E-\hbar\omega_0\right)\left(\bop\ket{\psi}\right),
    \end{align}
\end{subequations}
meaning that operating on the system $\ket{\psi}$ with $\bdag$ raises its energy by a quantum of $\hbar\omega_0$, while operating with $\bop$ lowers its energy by $\hbar\omega_0$.  This is akin to moving up and down the rung of a ladder, hence the name ``ladder operators''.  By definition, the lowest energy state is called the vacuum state $\ket{0}$, which satisfies $\hat{H}\left(\bop\ket{0}\right)=0$. The $n$th eigenstate is given by
\begin{equation}
    \ket{n} = \frac{1}{\sqrt{n!}}\bdagn{n}\ket{0}.
\end{equation}

The formulation of ladder operators is not limited to a mechanical oscillator. In general, it can be formulated from any system described by a Hamiltonian resembling Eq.~\eqref{eq:SHO_Ham} where the \emph{canonical} position $q$ and momentum $p$ take the roles of the real position and momentum considered above (see, for example Chap. 2 of Ref.~\cite{gerry_introductory_2005} for the treatment of electromagnetic waves).  In some cases, they are called the ``raising`` and ``lowering'' operators.  In some others, they are called the ``annihilation`` and ``creation`` operators.  In this paper, we shall use the term ``ladder operators'' as a reference to both operators, while the ``annihilation" and ``creation" operators refer to $\bop$ and $\bdag$, respectively.

The formulation can also be applied to an ensemble of many simple harmonic oscillators.  In this case, we have pairs of ladder operators $\left(\bop_j,\bdag_j\right)$, each associated with a single uncoupled oscillator in the ensemble indexed by $j$.  They satisfy the commutation relations
\begin{subequations}
\begin{align}
    \comm{\bop_j}{\bop_k} &= \comm{\bdag_j}{\bdag_k} = 0
    \label{eq:boson_comm_1},
    \\
    \comm{\bop_j}{\bdag_k} &= \delta_{jk},
    \label{eq:boson_comm_2}
\end{align}
\end{subequations}
where $\delta_{jk}$ is the Kronecker delta. 

The indices also appear in the occupation number representation~(see, for
example, Chap. 3 of Ref.~\cite{lancaster_quantum_2014}). Let us consider
$\left |{n_{0},n_{1},n_{2},\dots }\right \rangle $, where $n_{0}$ denotes
the number of particles in the quantum state $\psi _{0}$, $n_{1}$ is the
number of particles in the state $\psi _{1}$, etc. To decrease and increase the number of particles, we use the ladder operators
with the corresponding index: $\hat{b}_{j}$ and
$\hat{b}^{\dagger }_{j}$, respectively. Let us now consider $\left |{0,0}\right \rangle $ and try to increase each count by one. This
can be done in two ways:
\begin{equation}
    \bdag_1\bdag_0\ket{0,0}\propto\ket{1,1}
\end{equation}
or
\begin{equation}
    \bdag_0\bdag_1\ket{0,0}\propto\ket{1,1}.
\end{equation}
Since we end up in the same state, we must have
\begin{equation}
    \bdag_0\bdag_1=\lambda\bdag_1\bdag_0
\end{equation}
where $\lambda$ is a scalar.  By quantum statistics, in our three-dimensional world we have $\lambda=\pm 1$.  We can similarly consider other processes, such as adding one particle in the state $\psi_0$ and removing one in the state $\psi_1$.  In general, the $\lambda=1$ case corresponds to bosons and gives the commutation relations shown in Eqs.~\eqref{eq:boson_comm_1} and~\eqref{eq:boson_comm_2}.  Meanwhile, the $\lambda=-1$ case corresponds to fermions and gives the (anti)commutation relations
\begin{subequations}
\begin{align}
    \acomm{\hat{a}_j}{\hat{a}_k} &= \acomm{\hat{a}_j^\dagger}{\hat{a}_k^\dagger} = 0,
    \\
    \acomm{\hat{a}_j}{\hat{a}_k^\dagger} &= \delta_{jk},
\end{align}
\end{subequations}
where we have used $a$ instead of $b$ for fermionic ladder operators. 
 In the following, we focus only on the bosonic ladder operators and drop the adjective ``bosonic'' for brevity. 


\subsection{Normal ordering and Blasiak's formulae}\label{subsec:normal_ordering}

\begin{revision}Let us consider a monomial in the ladder operators, such as $\bdag\bop\bdag$ or $\bop^2\bdagn{2}$. A ladder monomial is said to be \emph{normal-ordered} if all creation operators are positioned to the left of all annihilation operators, i.e., a monomial of the form $\bdagn{p}\bop^q; p,q\in\mathbb{Z}$. Conventionally, the term ``normal ordering'' refers to Wick ordering, where a given monomial is replaced with an \emph{inequivalent} normal-ordered monomial containing the same number of creation and annihilation operators. It is represented by the ``double dot enclosure'', e.g. $:\mathrel{\bdag\bop\bdag}:=\bdagn{2}\bop$ and $:\mathrel{\bop^2\bdagn{2}}:=\bdagn{2}\bop^2$. In other words, the monomial has been reordered with $\comm{\bop_j}{\bdag_k}=0$ (instead of $\delta_{jk}$). In quantum field theory, Wick ordering is useful to avoid infinite self-energy and to develop the Wick's theorem (see, for example, Chaps. 4 and 18 in Ref.~\cite{lancaster_quantum_2014}). In quantum optics, it is useful to simplify equations in quadrature squeezing (see, for example, Chapter 7 in Ref.~\cite{gerry_introductory_2005}).

Throughout this paper, we define ``normal ordering'' as rewriting the monomial into an \emph{equivalent} expression using the commutation relations. The normal ordering of a ladder monomial $\hat{X}$ is denoted $\mathcal{N}\left(\hat{X}\right)$ herein, for example, 
\begin{equation}
    \mathcal{N}\left(\bdag\bop\bdag\right) = \bdag+\bdagn{2}\bop
\end{equation}
This is useful in quantum optics when dealing with coherent states of the harmonic oscillator, i.e. the state $\ket{\beta}$ satisfying $\bop\ket{\beta} = \beta\ket{\beta}, \beta\in\mathbb{Z}$, which implies that the matrix element $\matrixel{\beta}{\hat{X}}{\beta}$ of some monomial $\hat{X}$ in the coherent state basis can be straightforwardly evaluated if $\hat{X}$ is normal-ordered. It means if $g_\mathcal{N}\left(\bop,\bdag\right)$ is a normal-ordered polynomial in the ladder operators, then
\begin{equation}
    \matrixel{\beta}{g_\mathcal{N}\left(\bop,\bdag\right)}{\beta} = g_\mathcal{N}\left(\beta,\beta^*\right) .
\end{equation}
One interesting property is the optical equivalence theorem. Let $P(\beta)$ be the Glauber-Sudarshan $P$ function, one of the possible phase-space representations of a quantum system; then,
\begin{equation}\begin{split}
    \expval{g_\mathcal{N}\left(\bop,\bdag\right)} = \int \mathrm{d}^2\beta\ P\left(\beta\right)g_\mathcal{N}\left(\beta,\beta^*\right),
\end{split}\end{equation}
which means that we can conveniently obtain the expectation value of any ladder polynomial by normal-ordering the operator and replacing $\left(\bop,\bdag\right)$ by $\left(\beta,\beta^*\right)$, turning the expectation value integral into a simpler weighted average(see for example, Chap. 3 of Ref.~\cite{gerry_introductory_2005}).
\end{revision}

\begin{table*}[!t]
\centering

\caption{Some \texttt{SymPy} objects \begin{revision}relevant to\end{revision} \pybolano. \texttt{SymPy} is \inlinecode{ver. 1.13.3}.}

\begin{tabular}{p{0.33\linewidth}|>{\raggedright\arraybackslash}p{\dimexpr\linewidth-0.37\linewidth-2\tabcolsep-1.5\arrayrulewidth\relax}}
    \hline
    \hline
        \texttt{SymPy} object 
        & 
        Description
    \\
    \hline
    \hline
        \inlinecode{Number}
        &
        Atomic expression for numbers (\inlinecode{Integer}, \inlinecode{Float}, \inlinecode{Rational}). The \inlinecode{args} attribute is an empty tuple.
    \\
    \hline
        \inlinecode{Symbol}
        &
        Symbol expression. Useful for variables. The \inlinecode{args} attribute is a tuple containing the string input to the object's constructor.
    \\
    \hline
        \inlinecode{Add}
        &
        The sum object. The \inlinecode{args} attribute is a tuple of its summands.
    \\
    \hline
        \inlinecode{Mul}
        &
        The product object. The \inlinecode{args} attribute is a tuple of its factors.
    \\
    \hline
        \inlinecode{Pow}
        &
        The exponentiation object. The \inlinecode{args} attribute is a tuple \inlinecode{(b,e)} of its base and exponent.
    \\
    \hline
    \hline
\end{tabular}
\label{table_1}
\end{table*}

\begin{figure*}[!t]
    \centering
    \includegraphics[width=0.8\linewidth]{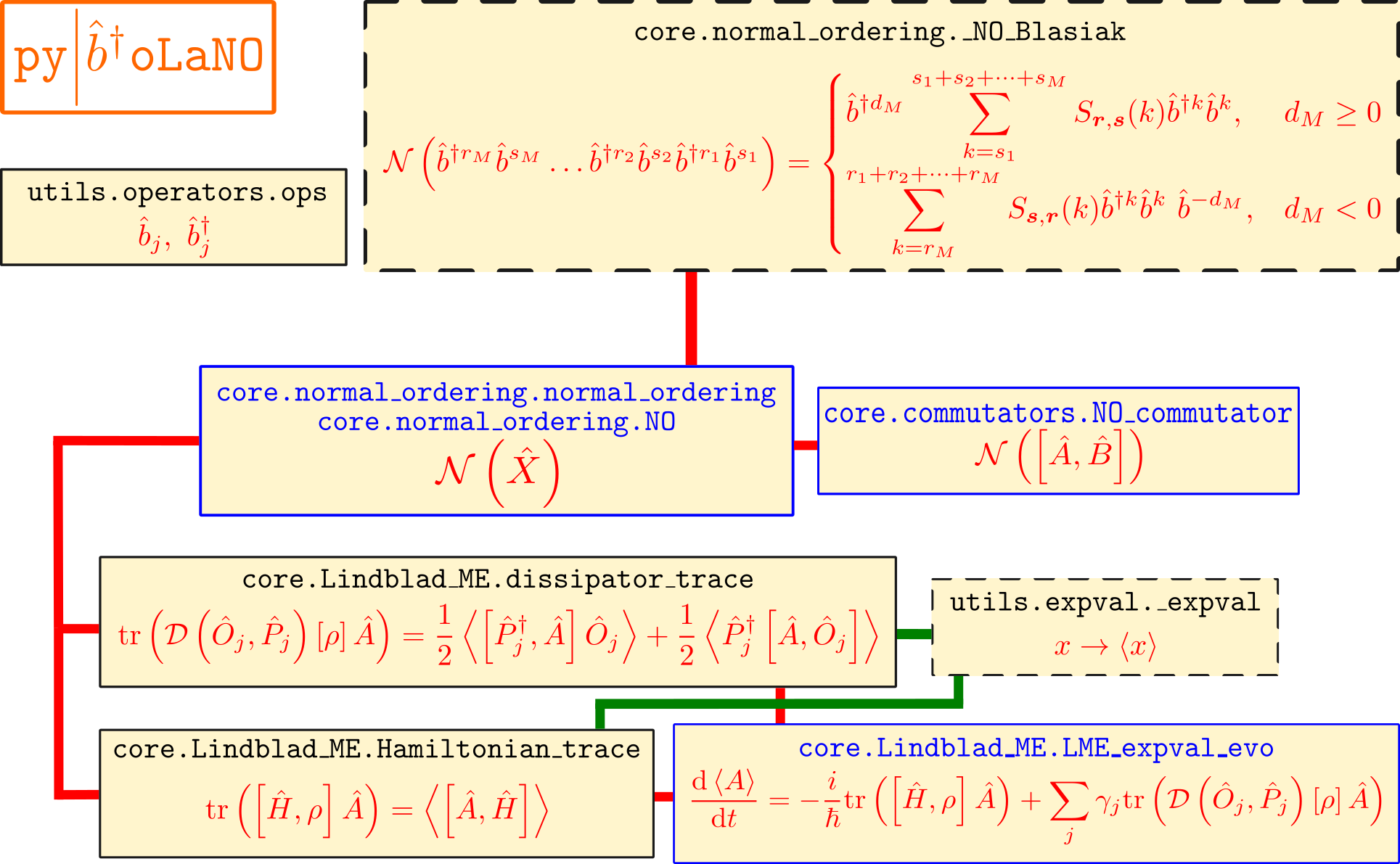}
    \caption{Core functionalities of \pybolano. The main functions are shown in blue, while the core utility functions are colored black. Meanwhile, the purple-colored \inlinecode{mp_config} is a \texttt{Python} variable. Red connectors show symbolic workflows, while green connectors show visual workflows. Dashed rectangles indicate functionalities that are not accessible by the user.}
    \label{fig_1}
\end{figure*}


A useful result is given by Blasiak~\cite{blasiak_combinatorics_2005, Mendez_2005} for the normal ordering of a ladder monomial. Let us now define
\begin{equation}
    \hat{X} = \bdagn{r_M}\bop^{s_M}\dots \bdagn{r_2}\bop^{s_2}\bdagn{r_1}\bop^{s_1} ,
\end{equation} 
$\bm{r}=\left(r_1,r_2,\dots,r_M\right)$, and $\bm{s}=\left(s_1,s_2,\dots,s_M\right)$.  Also, let 
\begin{equation}
    d_l = \sum_{m=1}^l\left(r_m-s_m\right)
\end{equation}
be the $l$th excess of creation operators in the monomial.  Given  the generalized Stirling numbers,
\begin{equation}\label{eq:S_rsk}
    S_{\bm{r},\bm{s}}(k) = \frac{1}{k!}\sum_{j=0}^k \binom{k}{j}\left(-1\right)^{k-j}\prod_{m=1}^M \left(d_{m-1}+j\right)_{s_m},
\end{equation}
where $\binom{a}{b}=\frac{a!}{b!(a-b)!}$ is the binomial coefficient and $(m)_n=\frac{m!}{(m-n)!}$ is the falling factorial, we can obtain the normal-ordered expression for $\hat{X}$ as 
\begin{equation}\label{eq:Blasiak}
    \mathcal{N}\left(\hat{X}\right) = 
    \begin{cases}\displaystyle
        \bdagn{d_M}\sum_{k=s_1}^{s_1+s_2+\dots+s_M} S_{\bm{r},\bm{s}}(k)\bdagn{k}\bop^k,
        &
        d_M \geq 0
        \\
        \displaystyle
        \sum_{k=r_M}^{r_1+r_2+\dots+r_M} S_{\overline{\bm{s}},\overline{\bm{r}}}(k)\bdagn{k}\bop^k\ \bop^{-d_M},
        &
        d_M<0
    \end{cases}
\end{equation}
where $\overline{\bm{r}}=\left(r_M,\dots,r_2,r_1\right)$ and $\overline{\bm{s}}=\left(s_M,\dots,s_2,s_1\right)$. 


\subsection{The Lindblad master equation}\label{subsec:LindbladME}

Normal ordering is practically used in open quantum systems to obtain expressions that are convenient to interpret and use in other calculations, e.g., the $Q$-parameter for photon statistics~\cite{gerry_introductory_2005}. A widely used formalism is the Lindblad master equation~\cite{schlosshauer_decoherence_2007, breuer_theory_2002}.  For a system described by the density matrix $\rho$, the evolution of the system is given by
\begin{equation}\label{eq:LME}
    \odv{\rho}{t} = -\frac{i}{\hbar}\comm{\hat{H}}{\rho} + \sum_j \gamma_j\dissip{\hat{O}_j, \hat{P}_j}.
\end{equation}
The Hamiltonian $\hat{H}$ describes the closed system dynamics, while the Lindblad dissipators 
\begin{equation}
    \dissip{\hat{O}_j, \hat{P}_j} = \hat{O}_j\rho\hat{P}_j^\dagger - \frac{1}{2}\acomm{\hat{P}_j^\dagger\hat{O}_j}{\rho}
\end{equation}
describe the open system dynamics. Each dissipator is defined by the operators $\hat{O}_j, \hat{P}_j$, which describe the open system process.  The multiplying scalar $\gamma_j\geq 0$ can be interpreted as the process rate. 

Theoretical treatment of open quantum systems often deals with calculating the evolution of expectation values for some physical quantities.  Let a quantity $A$ be represented by the operator $\hat{A}$.  Given the density matrix $\rho$, the expectation value of $A$ is given by $\trace{\rho \hat{A}}$.  Multiplying Eq.~\eqref{eq:LME} by $\hat{A}$ (from both sides of the equation), we have
\begin{equation}\begin{split}
    \odv{\expval{A}}{t} =& -\frac{i}{\hbar}\trace{\comm{\hat{H}}{\rho}\hat{A}} 
    \\
    &+ \sum_j \gamma_j \trace{\dissip{\hat{O}_j,\hat{P}_j}\hat{A}}
    \label{eq:LME_expval_evo}
\end{split}\end{equation}
We call the trace containing $\hat{H}$ the ``Hamiltonian trace'', and those containing $\mathcal{D}$ the ``dissipator traces''. For arbitrary Hamiltonian and dissipator operators, they are given by
\begin{align}
    \trace{\comm{\hat{H}}{\rho}\hat{A}} &= \expval{\comm{\hat{A}}{\hat{H}}}
    \label{eq:Ham_trace}
\end{align}
and
\begin{equation}\begin{split}\label{eq:dissip_trace}
    \trace{\dissip{\hat{O}_j,\hat{P}_j}\hat{A}} =& 
    \frac{1}{2}\expval{\comm{\hat{P}_j^\dagger}{\hat{A}}\hat{O}_j}
    \\
    &+\frac{1}{2}\expval{\hat{P}_j^\dagger\comm{\hat{A}}{\hat{O}_j}} .
\end{split}\end{equation}
These expressions become more cumbersome and error-prone to normal-order as the operators involved become more complex. This motivates us to automate the process with a computer program, \texttt{pyBoLaNO}, that we develop in this work.


\section{Package Description}\label{section:package_anatomy}

The core workflow of \pybolano is shown in Fig.~\ref{fig_1}. We start with a brief overview of the \texttt{SymPy} objects relevant to our package. Then, we describe the workflow of the package's main functions: \inlinecode{ops}, \inlinecode{normal_ordering}/\inlinecode{NO}, \inlinecode{NO_commutator}, and \inlinecode{LME_expval_evo}; as well as the multiprocessing configuration dictionary: \inlinecode{mp_config}.

\subsection{Relevant objects in \texttt{SymPy}}
\label{subsec_sympy_objects}

All algebraic expressions in \texttt{SymPy} are based on the \inlinecode{Expr} or \inlinecode{AtomicExpr} class. The former is used for expressions with arguments (composite expressions), such as \inlinecode{Symbol}. The latter is used for expressions without arguments (atomic expressions), such as \inlinecode{Number}. The tuple of arguments of an expression object can be accessed from its \inlinecode{args} attribute.  For \inlinecode{AtomicExpr}, an empty tuple is returned.

Table~\ref{table_1} briefly describes several \texttt{SymPy} objects relevant to our package. Each mathematical operation returns the corresponding operation class: \inlinecode{Add} for summation, \inlinecode{Mul} for multiplication, and \inlinecode{Pow} for exponentiation, to name a few. These are all composite objects and may have each other as arguments. \inlinecode{Add} and \inlinecode{Mul} may have any number of arguments. Meanwhile, the first argument of \inlinecode{Pow} is the base and the second argument is the exponent.  For example, given that \inlinecode{x=Symbol("x")}, then \inlinecode{x**2+2*x} is constructed as \inlinecode{Add(Pow(Symbol("x"),Integer(2)),Mul(Integer(2),} \inlinecode{Symbol("x")))}. This representation can be accessed using the \inlinecode{srepr} function. We can also check whether a given expression is contained using the \inlinecode{has} method.


\subsection{Constructing the ladder operators}

\begin{revision}The ladder operator objects are implemented as \inlinecode{BosonicAnnihilationOp} and \inlinecode{BosonicCreationOp}. Instead of initializing both objects, the package provides a convenient call through the function \inlinecode{ops}.\end{revision} By calling
\begin{minted}[bgcolor=red!15]{python}
ops(k)
\end{minted}
we can initialize both annihilation and creation operators with the same subscript. The subscript \inlinecode{k} is optional and can be \inlinecode{Symbol} or any \texttt{Python} object convertible to a string. This function preprocesses \inlinecode{k} into \inlinecode{Symbol} to be used as the subscript of ladder operators. 

\begin{revision2}
Since the objects inherit from \inlinecode{sympy.Expr} without any modifications to methods handling interactions between \texttt{SymPy} objects, they are compatible with other expression objects. In particular, they are compatible with the sigma and pi notations, implemented as \inlinecode{sympy.Sum} and \inlinecode{sympy.Product}, respectively. That is, the summation and product can be explicitly carried out, e.g. $\sum_{k=1}^3 \bop_k = \bop_1+\bop_2+\bop_3$ and $\prod_{k=1}^3 \bop_k = \bop_1\bop_2\bop_3$.
\end{revision2}

\subsection{Normal ordering}
\label{subsec_normal_ordering}

The normal ordering of a polynomial \inlinecode{q} in ladder operators is done by the \inlinecode{normal_ordering} function, with syntax
\begin{minted}[bgcolor=red!15]{python}
normal_ordering(q)
\end{minted}
We also offer a shorthand
\begin{minted}[bgcolor=red!15]{python}
NO(q)
\end{minted}
for the same purpose.  

Blasiak's formulae, given by Eq.~\eqref{eq:Blasiak}, are implemented in the \inlinecode{_NO_Blasiak} function, which evaluates the normal ordering of the given monomial in the ladder operators.  Extra algorithms are executed to handle multipartite descriptions.  The general workflow is described below.
\begin{enumerate}
    \item Make a list of summands of \inlinecode{q}.
    
    \item Make each summand of \inlinecode{q} into a list of factors, separated by the ladder operator subscripts. Scalars are put in a separate group.
    
    \item Normal-order each factor using \inlinecode{_NO_Blasiak}.
    
    \item Multiply the sums obtained by normal-ordering each factor to give the almost-normal-ordered summand of \inlinecode{q}. 
    
    \item Sum all the almost-normal-ordered summands of \inlinecode{q} to give the almost-normal-ordered \inlinecode{q}.
    
    \item Due to how \texttt{SymPy} handles noncommuting objects, the \inlinecode{Mul} of factors obtained in step 4 are yet to be normal-ordered. Instead, each term goes like $\dots \bdagn{a}_j\bop^b_j\bdagn{c}_{k\neq j}\bop^{d}_{k\neq j}\dots$. Do the final sorting as follows for each summand:
    \begin{enumerate}
        \item Collect scalars into one list.
        \item Collect creation operators as dictionary entries whose keys are their subscripts.
        \item Collect annihilation operators similarly.
        \item Sort both dictionaries by key, then return a list of their values. 
        \item Multiply all items in each list to give a \inlinecode{Mul} of scalars, creation operators, and annihilation operators.
        \item Multiply the three resulting \inlinecode{Mul} objects to give the fully normal-ordered summand.
    \end{enumerate}
    Sum all summands and return the resulting \inlinecode{Add}. 
\end{enumerate}


\subsection{Normal-ordered commutators}
\label{subsec_commutators}

The normal ordering of a commutator between two polynomials \inlinecode{A} and \inlinecode{B} in ladder operators can be evaluated using the \inlinecode{do_commutator} function.  The syntax is given by
\begin{minted}[bgcolor=red!15]{python}
do_commutator(A, B)
\end{minted}
This function simply calls \inlinecode{normal_ordering(A*B-B*A)}. 

\subsection{Expectation value evolution in the Lindblad master equation framework}

Equation~\eqref{eq:LME_expval_evo} can be evaluated by calling \inlinecode{LME_expval_evo}. The syntax is
\begin{minted}[bgcolor=red!15]{python}
LME_expval_evo(H, D, A, hbar_is_one)
\end{minted}
Here, \inlinecode{H} is the Hamiltonian of the system, written as a \texttt{SymPy} expression. Meanwhile, \inlinecode{D} is a list that specifies the dynamics of the open system.  Each element in \inlinecode{D} is a list \inlinecode{[gamma_j, O_j, P_j]} containing the process rate $\gamma_j$ and the operators $\hat{O}_j,\hat{P}_j$ defining the dissipator. If $\hat{O}_j=\hat{P}_j$, the user may omit the third element from the list. If no dissipation is present, the user can input an empty list, \inlinecode{D=[]}. The operator corresponding to the quantity whose expectation value evolution is calculated is input to \inlinecode{A}. Finally, \inlinecode{hbar_is_one} is a Boolean to set $\hbar=1$. This option is \inlinecode{True} by default. We note that \inlinecode{Symbol("hbar")} is used instead of \inlinecode{sympy.physics.quantum.hbar} to ensure that it behaves well with the package functionalities. The following happens when this function is called.
\begin{enumerate}
    \item Call \inlinecode{Hamiltonian_trace(H, A)} to calculate the Hamiltonian trace with Eq.~\eqref{eq:Ham_trace} in normal-ordered form. Add the result to the output.
    \item For each entry in \inlinecode{D}, call \inlinecode{dissipator_trace(O_j,P_j,A)} to calculate the dissipator trace given by Eq.~\eqref{eq:dissip_trace} in normal-ordered form. Multiply the result by the corresponding process rate \inlinecode{gamma_j} and add to the output.
    \item Return the output \inlinecode{Add}. 
\end{enumerate}
The output is generally a sum that contains \inlinecode{_expval} objects. The \inlinecode{_expval} class wraps the output operators inside bra-kets. The class inherits \texttt{SymPy}'s \inlinecode{Symbol} class and is used only for result visualization. 


\subsection{Mutiprocessing configurations}

Multiprocessing is used in \inlinecode{normal_ordering} to handle each summand of the input. The user may control the multiprocessing behavior by modifying the \inlinecode{mp_config} variable after importing the package. The syntax is
\begin{minted}[bgcolor=red!15]{python}
mp_config["enable"] = True
mp_config["num_cpus"] = os.cpu_count()
mp_config["min_num_args"] = 2
\end{minted}
where the values assigned above are the default values of the package. Here, \inlinecode{"enable"} specifies whether multiprocessing is enabled, \inlinecode{"num_cpus"} specifies the number of CPU threads to use, and \inlinecode{"min_num_args"} specifies the minimum number of summands mentioned above (arguments of the \inlinecode{Add} object). Any integer less than \inlinecode{2} will cause the package to set \inlinecode{mp_config["min_num_args"]=2}. 


\section{Some Use Cases of the Package}\label{section:examples}

The examples provided in this section are available in the package tutorial on our GitHub repository, which uses \texttt{Jupyter Notebook} to render the \LaTeX{} output. The following snippets show an exemplary setup of the package.
\begin{minted}[bgcolor=red!15]{python}
import pybolano as bl
bl.mpconfig["num_cpus"] = 4
\end{minted}

\begin{minted}{python}
import sympy as sm
latex = sm.latex # to print as LaTeX code.

b, bd = bl.ops()
b_1, bd_1 = bl.ops(1)
b_2, bd_2 = bl.ops(sm.Symbol("2")) # or any other symbols such as sm.Symbol("k") for a dummy index.

print(latex(
b_1
))

Out: b_{1}
\end{minted}
\begin{revision2}
\noindent \noindent\inlinecode{Output render:}
\begin{equation*}
    b_{1}
\end{equation*}
\end{revision2}

\subsection{\inlinecode{normal_ordering}}

\begin{minted}{python}
print(latex(
bl.normal_ordering(b * bd * b)
))

Out: b_{} + {b^\dagger_{}} b_{}^{2}
\end{minted}
\begin{revision2}
\inlinecode{Input render:}
\begin{equation*}
    \bop\bdag\bop
\end{equation*}
\noindent\noindent\inlinecode{Output render:}
\begin{equation*}
    b_{} + {b^\dagger_{}} b_{}^{2}
\end{equation*}
\end{revision2}

\noindent Multipartite input:
\begin{minted}{python}
print(latex(
bl.normal_ordering(b_2 * b_1 * bd_2**2 * bd_1)
))

Out: 2 {b^\dagger_{1}} {b^\dagger_{2}} b_{1} + {b^\dagger_{1}} {b^\dagger_{2}}^{2} b_{1} b_{2} + 2 {b^\dagger_{2}} + {b^\dagger_{2}}^{2} b_{2}
\end{minted}
\begin{revision2}
\noindent\inlinecode{Input render:}
\begin{equation*}
    \bop_2\bop_1\bdagn{2}_2\bdag_1
\end{equation*}
\noindent\inlinecode{Output render:}
\begin{equation*}
    2 {b^\dagger_{1}} {b^\dagger_{2}} b_{1} + {b^\dagger_{1}} {b^\dagger_{2}}^{2} b_{1} b_{2} + 2 {b^\dagger_{2}} + {b^\dagger_{2}}^{2} b_{2}
\end{equation*}
\end{revision2}

\noindent Polynomial input:
\begin{minted}{python}
print(latex(
bl.normal_ordering(b_1*bd_2 + 5*b_2**2*bd_1*b_1 + b_2)
))

Out: b_{2} + 5 {b^\dagger_{1}} b_{1} b_{2}^{2} + {b^\dagger_{2}} b_{1}
\end{minted}
\begin{revision2}
\inlinecode{Input render:}
\begin{equation*}
    \bop_1 \bdag_2 + 5\bop_2^2 \bdag_1\bop_1 + \bop_2
\end{equation*}
\noindent\inlinecode{Output render:}
\begin{equation*}
    b_{2} + 5 {b^\dagger_{1}} b_{1} b_{2}^{2} + {b^\dagger_{2}} b_{1}
\end{equation*}
\end{revision2}

\noindent Input with symbols:
\begin{minted}{python}
x = sm.Symbol("x")

print(latex(
bl.normal_ordering(x * b_1 * x**2 * bd_1**2)
))

Out: 2 x^{3} {b^\dagger_{1}} + x^{3} {b^\dagger_{1}}^{2} b_{1}
\end{minted}
\begin{revision2}
\inlinecode{Input render:}
\begin{equation*}
    x\bop_1 x^2 \bdagn{2}_1
\end{equation*}
\noindent\noindent\inlinecode{Output render:}
\begin{equation*}
    2 x^{3} {b^\dagger_{1}} + x^{3} {b^\dagger_{1}}^{2} b_{1}
\end{equation*}
\end{revision2}


\subsection{\inlinecode{NO_commutator}}

\begin{minted}{python}
A = bd*b
B = b

print(latex(
bl.NO_commutator(A, B)
))

Out: - b_{}     
\end{minted}
\begin{revision2}
\noindent\inlinecode{Input render:}
\begin{align*}
    \hat{A} &= \bdag\bop
    \\
    \hat{B} &= \bop
\end{align*}
\noindent\noindent\inlinecode{Output render:}
\begin{equation*}
    -\bop
\end{equation*}
\end{revision2}

\noindent Multipartite input:
\begin{minted}{python}
A = bd_1*bd_2
B = b_1*b_2

print(latex(
bl.NO_commutator(A, B)
))

Out: -1 - {b^\dagger_{1}} b_{1} - {b^\dagger_{2}} b_{2}
\end{minted}
\begin{revision2}
\noindent\inlinecode{Input render:}
\begin{align*}
    \hat{A} &= \bdag_1\bop_2
    \\
    \hat{B} &= \bop_1\bop_2
\end{align*}
\noindent\noindent\inlinecode{Output render:}
\begin{equation*}
    -1 - {b^\dagger_{1}} b_{1} - {b^\dagger_{2}} b_{2}
\end{equation*}
\end{revision2}

\noindent Polynomial input:
\begin{minted}{python}
A = b_1 + 2*b_2**2
B = bd_1**3 + 2*bd_2*b_2

print(latex(
bl.NO_commutator(A, B)
))

Out: 8 b_{2}^{2} + 3 {b^\dagger_{1}}^{2}
\end{minted}
\begin{revision2}
\noindent\inlinecode{Input render:}
\begin{align*}
    \hat{A} &= \bop_1+2\bop_2^2
    \\
    \hat{B} &= \bdagn{3}_1+2\bdag_2\bop_2
\end{align*}
\noindent\noindent\inlinecode{Output render:}
\begin{equation*}
    8 b_{2}^{2} + 3 {b^\dagger_{1}}^{2}
\end{equation*}
\end{revision2}

\newpage
\noindent Input with symbols:
\begin{minted}{python}
x = sm.Symbol("x")
A = x*b_1 
B = x**(0.5)*bd_1*b

print(latex(
bl.NO_commutator(A, B)
))

Out: x^{1.5} b_{}
\end{minted}
\begin{revision2}
\noindent\inlinecode{Input render:}
\begin{align*}
    \hat{A} &= x\bop_1
    \\
    \hat{B} &= x^{0.5}\bdag_1\bop
\end{align*}
\noindent\noindent\inlinecode{Output render:}
\begin{equation*}
    x^{1.5} b_{}
\end{equation*}
\end{revision2}


\subsection{\inlinecode{LME_expval_evo}}

Now we consider some evolution equations in the literature as a way to validate this package, highlighting its potential usage for describing various quantum systems.  The following examples also illustrate problems where it is desirable to compute the evolution of the expectation values of some observables for which \pybolano may be useful. 

\subsubsection{The quantum simple harmonic oscillator}

It is well-known that the evolution of the expected value of the annihilation operator $\bop$ in a simple harmonic Hamiltonian $\hat{H} = \hbar\omega_0\left(\bdag\bop+1/2\right)$ is given by
\begin{equation}
    \odv{\expval{\bop}}{t} = -i\omega_0\expval{\bop}
\end{equation}
Here is how it is obtained with \pybolano:
\begin{minted}{python}
hbar, omega_0 = sm.symbols(r"hbar omega_0")
b, bd = bl.ops()

H = hbar*omega_0*bd*b
D = []
A = b

print(latex(
bl.LME_expval_evo(H, D, A, hbar_is_one=False)
))

Out: \frac{d}{d t} {\left\langle b_{} \right\rangle} = - i \omega_{0} {\left\langle b_{} \right\rangle}
\end{minted}
\begin{revision2}
\noindent\noindent\inlinecode{Output render:}
\begin{equation*}
    \frac{d}{d t} {\left\langle b_{} \right\rangle} = - i \omega_{0} {\left\langle b_{} \right\rangle}
\end{equation*}
\end{revision2}
Additionally, we can show that the energy is conserved, i.e. that
\begin{equation}
    \odv{\expval{\bdag\bop}}{t} = 0
\end{equation}
\noindent Assuming the variables from the previous code block, we have
\begin{minted}{python}
A = bd*b

print(latex(
bl.LME_expval_evo(H, D, A)
))

Out: \frac{d}{d t} {\left\langle {b^\dagger_{}} b_{} \right\rangle} = 0
\end{minted}
\begin{revision2}
\noindent\noindent\inlinecode{Output render:}
\begin{equation*}
    \frac{d}{d t} {\left\langle b_{} \right\rangle} = - i \omega_{0} {\left\langle b_{} \right\rangle}
\end{equation*}
\end{revision2}

\subsubsection{The quantum Rayleigh oscillator}

A limit cycle can be defined as a stable oscillation to which the system is always attracted~\cite{pikovsky_synchronization_2001}. This is a consequence of nonlinearity and can be observed in nonlinear oscillators. The Rayleigh oscillator is one such exemplary model. We consider the evolution of the expected phase point $\expval{\bop}$ for a quantum Rayleigh oscillator, a nonlinear quantum oscillator exhibiting a quantum limit cycle. From Ref.~\cite{chia_relaxation_2020}, the equation of motion is specified by (with $\hbar=1$ and $\bop=\left[\hat{x}+i\hat{p}\right]/2$, cf. Eqs. [14--16] of Ref.~\cite{chia_relaxation_2020}):
\begin{equation}
\begin{split}
    \hat{H} &= \omega_0\bdag\bop + i\frac{\mu}{12}\left(\bdag\bop^3-\bdagn{3}\bop\right) 
    \\
    &\quad + i\frac{\mu}{24}\left(\bop^4-\bdagn{4}\right) - i\frac{\mu\left(q_0^2-1\right)}{4}\left(\bop^2-\bdagn{2}\right)
\end{split}
\end{equation}
and
\begin{subequations}
\begin{align}
    \gamma_1 &= \mu\left(q_0^2-1\right),\quad \hat{O}_1=\hat{P}_1 = \bdag 
    \\
    \gamma_2 &= \frac{3\mu}{4}, \quad \hat{O}_2 = \hat{P}_2=\bop^2
    \\
    \gamma_3 &= \mu, \quad \hat{O}_3=\hat{P}_3 = \bdag\bop - \frac{\bdagn{2}}{2}
\end{align}
\end{subequations}
where $\mu$ is the nonlinearity parameter and $q_0$ is one of the Rayleigh parameters. The evolution of $\expval{\bop}$ is governed by (cf. Eqs. [13, 17--25] of Ref.~\cite{chia_relaxation_2020})
\begin{equation}\label{eq:Rayleigh_expval_evo}
\begin{split}
    \odv{\expval{\bop}}{t} &= -i\omega_0\expval{\bop} + \frac{\mu}{2} \left(q_0^2-1\right) \left[\expval{\bop}+\expval{\bdag}\right] 
    \\
    &\quad -\frac{\mu}{6}\left[\expval{\bop^3}+\expval{\bdagn{3}}\right] - \frac{\mu}{2}\left[\expval{\bdag\bop^2}+\expval{\bdagn{2}\bop}\right]
\end{split}
\end{equation}
Here is how Eq.~\eqref{eq:Rayleigh_expval_evo} can be obtained with \pybolano:
\begin{minted}{python}
omega_0, mu, q_0 = sm.symbols(r"omega_0 mu q_0")
b, bd = bl.ops()

H = omega_0*bd*b \
    + sm.I*mu/12 * (bd*b**3 - bd**3*b) \
    + sm.I*mu/24 * (b**4-bd**4) \
    - sm.I*mu*(q_0**2-1)/4 * (b**2-bd**2)

D = [[mu*(q_0**2-1), bd],
     [3*mu/4, b**2],
     [mu, bd*b-bd**2/2]]

A = b

print(latex(
bl.LME_expval_evo(H, D, A)
))

Out: \frac{d}{d t} {\left\langle b_{} \right\rangle} = \frac{\mu q_{0}^{2} {\left\langle b_{} \right\rangle}}{2} + \frac{\mu q_{0}^{2} {\left\langle {b^\dagger_{}} \right\rangle}}{2} - \frac{\mu {\left\langle b_{} \right\rangle}}{2} - \frac{\mu {\left\langle b_{}^{3} \right\rangle}}{6} - \frac{\mu {\left\langle {b^\dagger_{}} \right\rangle}}{2} - \frac{\mu {\left\langle {b^\dagger_{}} b_{}^{2} \right\rangle}}{2} - \frac{\mu {\left\langle {b^\dagger_{}}^{2} b_{} \right\rangle}}{2} - \frac{\mu {\left\langle {b^\dagger_{}}^{3} \right\rangle}}{6} - i \omega_{0} {\left\langle b_{} \right\rangle}}
\end{minted}
\begin{revision2}
\noindent\inlinecode{Output render:}
\begin{align*}
\frac{d}{d t} {\left\langle b_{} \right\rangle} &= \frac{\mu q_{0}^{2} {\left\langle b_{} \right\rangle}}{2} + \frac{\mu q_{0}^{2} {\left\langle {b^\dagger_{}} \right\rangle}}{2} - \frac{\mu {\left\langle b_{} \right\rangle}}{2} - \frac{\mu {\left\langle b_{}^{3} \right\rangle}}{6} - \frac{\mu {\left\langle {b^\dagger_{}} \right\rangle}}{2} \\ &\quad - \frac{\mu {\left\langle {b^\dagger_{}} b_{}^{2} \right\rangle}}{2} - \frac{\mu {\left\langle {b^\dagger_{}}^{2} b_{} \right\rangle}}{2} - \frac{\mu {\left\langle {b^\dagger_{}}^{3} \right\rangle}}{6} - i \omega_{0} {\left\langle b_{} \right\rangle}
\end{align*}
\end{revision2}

\subsubsection{A bipartite quantum battery with quadratic driving}

A quantum battery is a quantum system that can store energy and whose energy can be harvested for useful work~\cite{downing_hyperbolic_2024}. Recently, Downing and Ukhtary~\cite{downing_hyperbolic_2024} proposed a setup for a bipartite quantum battery. The system contains a charger and a holder coupled to each other, with a short quadratic pulse driving the charger. The equation of motion is specified by (with $\hbar=1$, cf. Eqs. [1--6] of Ref.~\cite{downing_hyperbolic_2024})
\begin{equation}
\begin{split}
    \hat{H} &= \omega_c \bdag_c\bop_c + \omega_h\bdag_h\bop_h \\
    &\quad + g\left(\bdag_c\bop_h+\bdag_h\bop_c\right) + \frac{\Omega}{2}\delta(t)\left(\bdagn{2}_c+\bop^2_c\right)
\end{split}
\end{equation}
and
\begin{equation}
    \gamma_1 = \gamma, \quad \hat{O}_1 =\hat{O}_1= \bop_c
\end{equation}
where $g$ is the coupling strength, $\Omega$ is the pulse strength, and $\delta(t)$ is the Dirac delta function. The quantities $\expval{\bdag_c\bop_c}$ and $\expval{\bdag_h\bop_h}$ are proportional to the energy of the charger and the holder, respectively. After $t=0$, the driving is off, and they evolve according to (cf. Eq.~[12] of Ref.~\cite{downing_hyperbolic_2024})
\begin{subequations}
\begin{align}
\odv{\expval{\bdag_c\bop_c}}{t} &= -\gamma \expval{\bdag_c\bop_c} - ig\left[\expval{\bdag_c\bop_h}-\expval{\bdag_h\bop_c}\right]  
\\
\odv{\expval{\bdag_h\bop_h}}{t} &= ig\left[\expval{\bdag_c\bop_h} - \expval{\bdag_h\bop_c}\right]
\end{align}
\end{subequations}
Here are the same equations obtained using our package:
\begin{minted}{python}
omega_c, omega_h, g, gamma = \
    sm.symbols(r"omega_c omega_h g gamma")
b_c, bd_c = bl.ops("c")
b_h, bd_h = bl.ops("h")

H = omega_c * bd_c*b_c \
    + omega_h * bd_h*b_h \
    + g*(bd_c*b_h + bd_h*b_c)

D = [[gamma, b_c]]

A = bd_c*b_c

print(latex(
bl.LME_expval_evo(H,D,A)
))

Out: \frac{d}{d t} {\left\langle {b^\dagger_{\mathtt{\text{c}}}} b_{\mathtt{\text{c}}} \right\rangle} = - i g {\left\langle {b^\dagger_{\mathtt{\text{c}}}} b_{\mathtt{\text{h}}} \right\rangle} + i g {\left\langle {b^\dagger_{\mathtt{\text{h}}}} b_{\mathtt{\text{c}}} \right\rangle} - \gamma {\left\langle {b^\dagger_{\mathtt{\text{c}}}} b_{\mathtt{\text{c}}} \right\rangle}
\end{minted}
\begin{revision2}
\noindent\inlinecode{Output render:}
\begin{align*}
\frac{d}{d t} {\left\langle {b^\dagger_{\mathtt{\text{c}}}} b_{\mathtt{\text{c}}} \right\rangle} = - i g {\left\langle {b^\dagger_{\mathtt{\text{c}}}} b_{\mathtt{\text{h}}} \right\rangle} + i g {\left\langle {b^\dagger_{\mathtt{\text{h}}}} b_{\mathtt{\text{c}}} \right\rangle} - \gamma {\left\langle {b^\dagger_{\mathtt{\text{c}}}} b_{\mathtt{\text{c}}} \right\rangle}
\end{align*}
\end{revision2}
\noindent Assuming the variables from the previous block, 
\begin{minted}{python}
A = bd_h*b_h

print(latex(
bl.LME_expval_evo(H,D,A)
))

Out: \frac{d}{d t} {\left\langle {b^\dagger_{\mathtt{\text{h}}}} b_{\mathtt{\text{h}}} \right\rangle} = i g {\left\langle {b^\dagger_{\mathtt{\text{c}}}} b_{\mathtt{\text{h}}} \right\rangle} - i g {\left\langle {b^\dagger_{\mathtt{\text{h}}}} b_{\mathtt{\text{c}}} \right\rangle}
\end{minted}
\begin{revision2}
\noindent\inlinecode{Output render:}
\begin{align*}
\frac{d}{d t} {\left\langle {b^\dagger_{\mathtt{\text{h}}}} b_{\mathtt{\text{h}}} \right\rangle} = i g {\left\langle {b^\dagger_{\mathtt{\text{c}}}} b_{\mathtt{\text{h}}} \right\rangle} - i g {\left\langle {b^\dagger_{\mathtt{\text{h}}}} b_{\mathtt{\text{c}}} \right\rangle}
\end{align*}
\end{revision2}


\subsubsection{A $\mathcal{P}\mathcal{T}$-symmetric trimer of harmonic oscillators}

Bender and Boettcher~\cite{Bender1998} show that a Hamiltonian does not need to be Hermitian to have a real, nonnegative spectrum. The parity-time ($\mathcal{P}\mathcal{T}$) symmetry can be viewed as a generalization to the hermicity condition---this validates the non-Hermitian quantum mechanics formulation. In non-Hermitian quantum mechanics, exceptional points (EPs) mark the crossover between the unbroken and broken $\mathcal{P}\mathcal{T}$ phase. Exceptional points are useful. For example, at its EPs, a system may gain enhanced sensitivity, making it desirable for sensing applications~\cite{Wiersig2020}. As another example, it is observed that the fluorescence rates of certain single-photon sources reach their quantum limit upon crossing an EP~\cite{Zhou2024}. 

Downing and Saroka~\cite{Downing2021} formulate a simple model of short oligomer chains of harmonic oscillators with a $\mathcal{P}\mathcal{T}$-symmetric Hamiltonian in the Lindblad master equation framework, showing the emergence of EPs. We consider a trimer system whose dynamics is specified by ($\hbar=1$, cf. Eqs. [4, 5a, 5b, 10] of Ref.~\cite{Downing2021})
\begin{equation}
\begin{split}
    \hat{H} &= \left(\omega_0+i\frac{\kappa}{2}\right)\bdag_1\bop_1 + \omega_0\bdag_2\bop_2 + \left(\omega_0-i\frac{\kappa}{2}\right)\bdag_3\bop_3
    \\
    &\quad + g\left(\bdag_1\bop_2+\bdag_2\bop_3+\mathrm{h.c.}\right)
\end{split}
\end{equation}
and
\begin{subequations}
\begin{align}
    \gamma_1&=\gamma_1,\quad \hat{O}_1=\hat{P}_1=\bop_1
    \\
    \gamma_2&=\gamma_2,\quad \hat{O}_2=\hat{P}_2=\bop_2
    \\
    \gamma_3&=\gamma_3,\quad \hat{O}_3=\hat{P}_3=\bop_3
    \\
    \gamma_4&=p_1,\quad \hat{O}_4=\hat{P}_4=\bdag_1
    \\
    \gamma_5&=p_2,\quad \hat{O}_5=\hat{P}_5=\bdag_2
    \\
    \gamma_6&=p_3,\quad \hat{O}_6=\hat{P}_6=\bdag_3
\end{align}
\end{subequations}
where $\omega_0$ is the natural frequency of all oscillators, $g$ is the coupling strength, and $\kappa$ specifies both the gain rate of oscillator $1$ and loss rate of oscillator $3$. Meanwhile, $\gamma_k$ and $p_k$, $k=1,2,3$, specify the gain and loss rates from incoherent processes for the oscillator $k$. The non-Hermitian parts of $\hat{H}$ are consequences of these processes. The evolution of $\expval{\bdag_k\bop_k}$ are given by (cf. Eqs.[9, 14--19] of Ref.~\cite{Downing2021})
\begin{subequations}
\begin{align}
    \begin{split}
        \odv{\expval{\bdag_1\bop_1}}{t}
        &= p_1-\left(\gamma_1-p_1\right)\expval{\bdag_1\bop_1}
        \\ 
        &\quad -ig\expval{\bdag_1\bop_2}+ig\expval{\bdag_2\bop_1}
    \end{split}
    \\
    \begin{split}
    \odv{\expval{\bdag_2\bop_2}}{t} 
    &= 
    p_2 - \left(\gamma_2-p_2\right)\expval{\bdag_2\bop_2}
    \\ 
    &\quad +ig\expval{\bdag_1\bop_2}  -ig \expval{\bdag_2\bop_1}
    \\ 
    &\quad - ig\expval{\bdag_2\bop_3} + ig\expval{\bdag_3\bop_2}
    \end{split}
    \\
    \begin{split}
    \odv{\expval{\bdag_3\bop_3}}{t} 
    &= 
    p_3 - \left(\gamma_3-p_3\right) \expval{\bdag_3\bop_3}
    \\ 
    &\quad +ig\expval{\bdag_2\bop_3} - ig\expval{\bdag_3\bop_2}
    \end{split}
\end{align}
\end{subequations}
Here are the equations obtained using \pybolano:
\begin{minted}{python}
omega_0, kappa = sm.symbols("omega_0 kappa")
gamma_1, gamma_2, gamma_3 = sm.symbols("gamma_1 gamma_2 gamma_3")
p_1, p_2, p_3 = sm.symbols("p_1 p_2 p_3")

b_1, bd_1 = bl.ops(1)
b_2, bd_2 = bl.ops(2)
b_3, bd_3 = bl.ops(3)

H = (omega_0 + sm.I*kappa/2)*bd_1*b_1 \
    + omega_0*bd_2*b_2 \
    + (omega_0 - sm.I*kappa/2)*bd_3*b_3 \
    + g*(bd_1*b_2+bd_2*b_1 + bd_2*b_3 + bd_3*b_2)
    
D = [[gamma_1, b_1],
     [gamma_2, b_2],
     [gamma_3, b_3],
     [p_1, bd_1],
     [p_2, bd_2],
     [p_3, bd_3]]

A = bd_1*b_1

print(latex(
bl.LME_expval_evo(H, D, A)
))

Out: \frac{d}{d t} {\left\langle {b^\dagger_{1}} b_{1} \right\rangle} = - i g {\left\langle {b^\dagger_{1}} b_{2} \right\rangle} + i g {\left\langle {b^\dagger_{2}} b_{1} \right\rangle} - \gamma_{1} {\left\langle {b^\dagger_{1}} b_{1} \right\rangle} + p_{1} {\left\langle {b^\dagger_{1}} b_{1} \right\rangle} + p_{1}
\end{minted}
\begin{revision2}
\noindent\inlinecode{Output render:}
\begin{align*}
\frac{d}{d t} {\left\langle {b^\dagger_{1}} b_{1} \right\rangle} &= - i g {\left\langle {b^\dagger_{1}} b_{2} \right\rangle} + i g {\left\langle {b^\dagger_{2}} b_{1} \right\rangle} \\&\quad - \gamma_{1} {\left\langle {b^\dagger_{1}} b_{1} \right\rangle} + p_{1} {\left\langle {b^\dagger_{1}} b_{1} \right\rangle} + p_{1}
\end{align*}
\end{revision2}

\noindent \noindent Assuming the variables from the previous block,
\begin{minted}{python}
A = bd_2*b_2

print(latex(
bl.LME_expval_evo(H, D, A)
))

Out: \frac{d}{d t} {\left\langle {b^\dagger_{2}} b_{2} \right\rangle} = i g {\left\langle {b^\dagger_{1}} b_{2} \right\rangle} - i g {\left\langle {b^\dagger_{2}} b_{1} \right\rangle} - i g {\left\langle {b^\dagger_{2}} b_{3} \right\rangle} + i g {\left\langle {b^\dagger_{3}} b_{2} \right\rangle} - \gamma_{2} {\left\langle {b^\dagger_{2}} b_{2} \right\rangle} + p_{2} {\left\langle {b^\dagger_{2}} b_{2} \right\rangle} + p_{2}
\end{minted}
\begin{revision2}
\noindent\inlinecode{Output render:}
\begin{align*}
\frac{d}{d t} {\left\langle {b^\dagger_{2}} b_{2} \right\rangle} &= i g {\left\langle {b^\dagger_{1}} b_{2} \right\rangle} - i g {\left\langle {b^\dagger_{2}} b_{1} \right\rangle} \\ &\quad - i g {\left\langle {b^\dagger_{2}} b_{3} \right\rangle} + i g {\left\langle {b^\dagger_{3}} b_{2} \right\rangle} 
\\&\quad - \gamma_{2} {\left\langle {b^\dagger_{2}} b_{2} \right\rangle} + p_{2} {\left\langle {b^\dagger_{2}} b_{2} \right\rangle} + p_{2}
\end{align*}
\end{revision2}
\noindent and
\begin{minted}{python}
A = bd_3*b_3

print(latex(
bl.LME_expval_evo(H, D, A)
))

Out: \frac{d}{d t} {\left\langle {b^\dagger_{3}} b_{3} \right\rangle} = i g {\left\langle {b^\dagger_{2}} b_{3} \right\rangle} - i g {\left\langle {b^\dagger_{3}} b_{2} \right\rangle} - \gamma_{3} {\left\langle {b^\dagger_{3}} b_{3} \right\rangle} + p_{3} {\left\langle {b^\dagger_{3}} b_{3} \right\rangle} + p_{3}
\end{minted}
\begin{revision2}
\noindent\inlinecode{Output render:}
\begin{align*}
\frac{d}{d t} {\left\langle {b^\dagger_{3}} b_{3} \right\rangle} &= i g {\left\langle {b^\dagger_{2}} b_{3} \right\rangle} - i g {\left\langle {b^\dagger_{3}} b_{2} \right\rangle} 
\\&\quad - \gamma_{3} {\left\langle {b^\dagger_{3}} b_{3} \right\rangle} + p_{3} {\left\langle {b^\dagger_{3}} b_{3} \right\rangle} + p_{3}
\end{align*}
\end{revision2}


\subsubsection{A pair of nonreciprocal driven-dissipative quantum resonators}

So far, we have considered the cases where $\hat{O}_j=\hat{P}_j$.  For the last example, we will discuss a system in which this is not the case. 

The notion of reciprocity, i.e., when the interaction between two objects is identical if they are swapped, is encountered in various topics in physics, such as Newton's third law of motion, electromagnetic reciprocity~\cite{Caloz2018}, and scattering reciprocity~\cite{Dek2012}. The lack of reciprocity (in other words: nonreciprocity) can lead to interesting phenomena and practical advantages~\cite{Downing2022}. A physical model that gives rise to nonreciprocity in open quantum systems is given by Downing and Sturges~\cite{Downing2022}. The model consists of a pair of driven-dissipative quantum resonators, where the asymmetry arises from the relative phase difference between the coherent and incoherent couplings. The dynamics is described within the Lindblad master equation framework, with ($\hbar=1$, cf. Eqs. [1, 2] of Ref.~\cite{Downing2022})
\begin{equation}
    \hat{H} = \Delta\sum_{k=1,2}\bdag_k\bop_k + \Omega\left(\bop_1+\bdag_1\right) + g\left(e^{i\theta}\bdag_1\bop_2 + e^{-i\theta}\bdag_2\bop_1\right) 
\end{equation}
and
\begin{subequations}
\begin{align}
    \gamma_1 &= \gamma, \quad \hat{O}_1=\hat{P}_1=\bop_1
    \\
    \gamma_2 &= \gamma, \quad \hat{O}_2=\hat{P}_2=\bop_2
    \\
    \gamma_3 &= \Gamma e^{i\phi}, \quad \hat{O}_3=\bop_2,\quad \hat{P}_3=\bop_1
    \\
    \gamma_4 &= \Gamma e^{-i\phi}, \quad \hat{O}_4 = \bop_1,\quad \hat{P}_4 = \bop_2
\end{align}
\end{subequations}
in the rotating reference frame of the laser driving oscillator $1$. Here, $\Delta$ is the detuning of the oscillators with respect to the laser, $\Omega$ is the driving strength, $g$ is the (coherent) coupling strength, and $\theta$ is the coupling phase.  The third and fourth dissipators describe the incoherent coupling between the two oscillators sharing the same bath, whose rate is taken to be complex with amplitude $0\leq \Gamma\leq \gamma$ and phase $\phi$.  The quantities $\expval{\bop_1}$ and $\expval{\bop_2}$ evolve according to (cf. Eqs. [3, 4] of Ref.~\cite{Downing2022})
\begin{subequations}
\begin{align}
        \odv{\expval{\bop_1}}{t} &=
    -i\left(\Delta -i \frac{\gamma}{2}\right)\expval{\bop_1}-\left(ige^{i\theta}+\frac{\Gamma e^{i\phi}}{2}\right)\expval{\bop_2}-i\Omega \label{eq:ex5_1}
    \\
    \odv{\expval{\bop_2}}{t} &=
    -\left(ige^{-i\theta}+\frac{\Gamma e^{-i\phi}}{2}\right)\expval{\bop_1}-i\left(\Delta-i\frac{\gamma}{2}\right)\expval{\bop_2}\label{eq:ex5_2}
\end{align}
\end{subequations}
Here are Eqs.~\eqref{eq:ex5_1} and~\eqref{eq:ex5_2} obtained using the package:
\begin{minted}{python}
Delta, Omega, g, theta = sm.symbols("Delta Omega g theta")
gamma, Gamma, phi = sm.symbols("gamma Gamma phi")

b_1, bd_1 = bl.ops(1)
b_2, bd_2 = bl.ops(2)

H = Delta*(bd_1*b_1 + bd_2*b_2) \
    + Omega*(b_1+bd_1) \
    + g*(sm.E**(sm.I*theta)*bd_1*b_2 + sm.E**(-sm.I*theta)*bd_2*b_1)

D = [[gamma, b_1],
     [gamma, b_2],
     [Gamma*sm.E**(sm.I*phi), b_2, b_1],
     [Gamma*sm.E**(-sm.I*phi), b_1, b_2]]

A = b_1

print(latex(
bl.LME_expval_evo(H, D, A)
))

Out: \frac{d}{d t} {\left\langle b_{1} \right\rangle} = - i \Delta {\left\langle b_{1} \right\rangle} - \frac{\Gamma {\left\langle b_{2} \right\rangle} e^{i \phi}}{2} - i \Omega - i g {\left\langle b_{2} \right\rangle} e^{i \theta} - \frac{\gamma {\left\langle b_{1} \right\rangle}}{2}
\end{minted}
\begin{revision2}
\noindent\inlinecode{Output render:}
\begin{align*}
\frac{d}{d t} {\left\langle b_{1} \right\rangle} = - i \Delta {\left\langle b_{1} \right\rangle} - \frac{\Gamma {\left\langle b_{2} \right\rangle} e^{i \phi}}{2} - i \Omega - i g {\left\langle b_{2} \right\rangle} e^{i \theta} - \frac{\gamma {\left\langle b_{1} \right\rangle}}{2}
\end{align*}
\end{revision2}
\newpage
\noindent Assuming variables from the previous block,
\begin{minted}{python}
A = b_2

print(latex(
bl.LME_expval_evo(H, D, A)
))

Out: \frac{d}{d t} {\left\langle b_{2} \right\rangle} = - i \Delta {\left\langle b_{2} \right\rangle} - \frac{\Gamma {\left\langle b_{1} \right\rangle} e^{- i \phi}}{2} - i g {\left\langle b_{1} \right\rangle} e^{- i \theta} - \frac{\gamma {\left\langle b_{2} \right\rangle}}{2}
\end{minted}
\begin{revision2}
\noindent\inlinecode{Output render:}
\begin{align*}
\frac{d}{d t} {\left\langle b_{2} \right\rangle} = - i \Delta {\left\langle b_{2} \right\rangle} - \frac{\Gamma {\left\langle b_{1} \right\rangle} e^{- i \phi}}{2} - i g {\left\langle b_{1} \right\rangle} e^{- i \theta} - \frac{\gamma {\left\langle b_{2} \right\rangle}}{2}
\end{align*}
\end{revision2}

\section{Performance}\label{section:performance}

\begin{revision} Since multiprocessing is implemented for each summand in the input expression, the speedup gained is approximately linear (the evaluation of each summand may take different durations). Meanwhile, the evaluation of a single monomial is optimal because explicit formulae are used. Considering Eq.~\eqref{eq:Blasiak}, we see that the number of terms in the normal-ordered equivalent of the input expression depends on the powers $\left\{s_k\right\}$ of $\bop$ for nonnegative excess (more $\bdag$ than $\bop$), and $\left\{r_k\right\}$ for negative excess. Furthermore, from Eq.~\eqref{eq:S_rsk} we can see that the calculation of the generalized Stirling number $S_{\bm{r},\bm{s}}(k)$ becomes more costly as we have more $\bop$ or $\bdag$ (more indices to sum over) in the expression and for larger $M$ (more indices to multiply over for the given term in the sum).

In comparison to our package, the original \texttt{SymPy} implements what we call the ``recursive flatten-and-swap`` algorithm, which can be roughly described as follows:
\begin{enumerate}
    \item For each summand in the input expression \inlinecode{expr}, make a list of its factors where \inlinecode{Pow} objects are flattened into multiple ladder operator objects. 
    \item Iterate through each item in the list. If the sequence $(\bop_j,\bdag_k)$ is found, swap their position using the commutation relations.
    \item Multiply together the resulting factors to generally get an \inlinecode{Add} object.
    \item Repeat steps 1--3 recursively until the resulting expression is not an \inlinecode{Add}, in which case it is added as an output summand. 
    \item Add together output summands to get the normal-ordered equivalent of \inlinecode{expr}.
\end{enumerate}
The algorithm creates a recursion tree where factor-listing and operator swapping are done at all nodes except the leaves. This quickly increases the computational cost as the input expression becomes more complex. 

The algorithm implemented by \texttt{pyBoLaNO}, on the other hand, provides a significant speedup over the conventional implementation. This is evident in our benchmark results as shown in Fig.~\ref{fig_2}. Our benchmarks are done in one node of our homebuilt computer cluster, Quasi Lab, running an \texttt{Intel i9-13900K} with $64\ \mathrm{GB}$ of RAM on a \texttt{Debian GNU/Linux 12 (bookworm) x86\_64} operating system. 

In our first benchmark, we time the normal ordering of $1000$ random monomials containing $10$ ladder operators of $2$ subsystems. The result in Fig.~\ref{fig_2}(a) shows that our algorithm can normal-order the given input about an order of magnitude faster than that of \texttt{SymPy}. There are cases where \texttt{SymPy} is faster.  Upon further inspection, we find that these are the cases where the input monomial is already normal-ordered. Our algorithm does not bypass the normal ordering process if the input is normal-ordered; adding this feature would introduce additional computational costs for other inputs. On the other hand, \texttt{SymPy} does this as a consequence of the algorithm it implements. The different spreads of the execution times for both algorithms are characteristic of them for the inputs used and are irrelevant to this discussion. 

\begin{figure}[!t]
    \centering
    \includegraphics[width=0.98\linewidth]{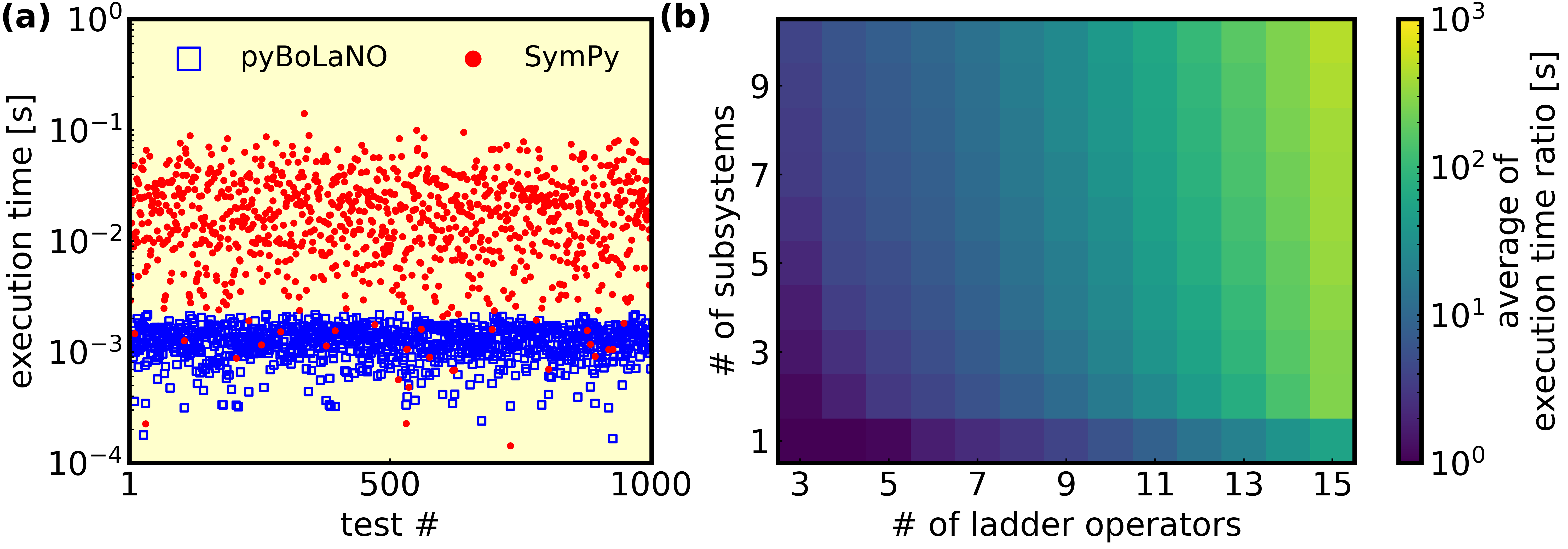}
    \caption{\begin{revision}Benchmarks of \texttt{pyBoLaNO}'s normal ordering algorithm against \texttt{sympy.physics.operatorordering.normal\_ordererd\_form} involving random ladder operator monomials. \textbf{Benchmark (a):} execution time of the normal ordering of $1000$ monomials containing $10$ ladder operators of $2$ subsystems. \textbf{Benchmark (b):} Average of the ratio between the average execution time of the normal ordering of $1000$ monomials by \texttt{SymPy} and \texttt{pyBoLaNO}, for varying numbers of ladder operators and subsystems. Note: these results may not be exactly reproducible due to the random nature of the benchmarks.\end{revision}}
    \label{fig_2}
\end{figure}

In our second benchmark, we take the average of the ratio between \texttt{SymPy}'s and our package's execution time over $1000$ random ladder monomials, observing how the value varies with the number of ladder operators in the input and the number of subsystems involved. Figure~\ref{fig_2} (b) shows a log-scale tile plot of our result. Evidently, as the numbers of ladder operators and subsystems increase, \texttt{SymPy}'s average execution time quickly grows to become multiple orders of magnitude above our package. This illustrates the superiority of the explicit formula that we implement.  We finally note that our package's multiprocessing does not help it perform better in these benchmarks since the input is a monomial. 

\end{revision}


\section{Conclusion}\label{section:conclusion}

We have developed a symbolic package \pybolano for the normal ordering of polynomials in bosonic ladder operators utilizing Blasiak's formulae. By extension, it allows the user to obtain the normal-ordered equivalent of a commutator between two polynomials in bosonic ladder operators, as well as the normal-ordered expectation value evolution equation for a system described in the Lindblad master equation framework. We described the package workflow in detail and provided the syntaxes for the core functionalities. We have exhibited some examples of use by taking recent results from the literature, which also serve to validate the package. We have discussed the computational cost of the normal ordering algorithm, \begin{revision}showing its superiority against the conventional implementation of \texttt{SymPy}.\end{revision} This package is aimed at quantum physics theorists who desire fast, error-free normal ordering for bosonic ladder algebra, in particular when dealing with expectation value evolution in the Lindblad master equation framework. We welcome suggestions and constructive criticism to improve this package through its public repository. 


\medskip
\noindent\textbf{Declaration of competing interest}
\medskip

The authors declare that they have no known competing financial interests or personal relationships that could have appeared to influence the work reported in this paper.

\medskip
\noindent\textbf{Data availability}
\medskip

The package's source code and instructions for installation are publicly available at \href{https://github.com/hendry24/pyBoLaNO}{\seqsplit{https://github.com/hendry24/pyBoLaNO}}. The code used for Section~\ref{section:examples} is compiled into a \texttt{Jupyter Notebook} available at \href{https://github.com/hendry24/pyBoLaNO/blob/main/tutorial.ipynb}{\seqsplit{https://github.com/hendry24/pyBoLaNO/blob/main/tutorial.ipynb}}. \begin{revision}The code used for Section~\ref{section:performance} is compiled into a \texttt{Jupyter Notebook} available at \href{https://github.com/hendry24/pyBoLaNO/blob/main/benchmarks.ipynb}{\seqsplit{https://github.com/hendry24/pyBoLaNO/blob/main/benchmarks.ipynb}}.\end{revision}

\medskip
\noindent\textbf{Acknowledgments}
\medskip

H.M.L., under the supervision of A.R.T.N., is supported by a research assistantship from the BRIN Directorate for Talent Management with
Grant No. 8/HK/II/2024.  All authors acknowledge Quasi Lab and Mahameru BRIN for their mini-cluster and HPC facilities. 


\bibliographystyle{elsarticle-num}
\bibliography{preprintV2/pybolanoPreprintV2}

\begin{thebibliography}{10}
\expandafter\ifx\csname url\endcsname\relax
  \def\url#1{\texttt{#1}}\fi
\expandafter\ifx\csname urlprefix\endcsname\relax\def\urlprefix{URL }\fi
\expandafter\ifx\csname href\endcsname\relax
  \def\href#1#2{#2} \def\path#1{#1}\fi

\bibitem{griffiths_introduction_2018}
D.~J. Griffiths, D.~F. Schroeter, Introduction to quantum mechanics, 3rd Edition, Cambridge University Press, 2018.

\bibitem{lancaster_quantum_2014}
T.~Lancaster, S.~Blundell, Quantum field theory for the gifted amateur, 1st Edition, Oxford Univ. Press, 2014.

\bibitem{gerry_introductory_2005}
C.~C. Gerry, P.~L. Knight, Introductory quantum optics, Cambridge University Press, 2005.

\bibitem{fox2006quantum}
M.~Fox, Quantum Optics: An Introduction, Oxford Master Series in Physics, OUP Oxford, 2006.

\bibitem{chia_relaxation_2020}
A.~Chia, L.~C. Kwek, C.~Noh, Relaxation oscillations and frequency entrainment in quantum mechanics, Phys. Rev. E 102~(4) (2020) 042213.

\bibitem{Shen2023}
Y.~Shen, W.-K. Mok, C.~Noh, A.~Q. Liu, L.-C. Kwek, W.~Fan, A.~Chia, Quantum synchronization effects induced by strong nonlinearities, Phys. Rev. A 107~(5) (2023) 053713.

\bibitem{downing_hyperbolic_2024}
C.~A. Downing, M.~S. Ukhtary, Hyperbolic enhancement of a quantum battery, Phys. Rev. A 109~(5) (2024) 052206.

\bibitem{Krimer2019}
D.~O. Krimer, M.~Zens, S.~Rotter, Critical phenomena and nonlinear dynamics in a spin ensemble strongly coupled to a cavity. {I}. semiclassical approach, Phys. Rev. A 100~(1) (2019) 013855.

\bibitem{Zens2019}
M.~Zens, D.~O. Krimer, S.~Rotter, Critical phenomena and nonlinear dynamics in a spin ensemble strongly coupled to a cavity. {II}. semiclassical-to-quantum boundary, Phys. Rev. A 100~(1) (2019) 013856.

\bibitem{Ahmadi2024}
B.~Ahmadi, P.~Mazurek, P.~Horodecki, S.~Barzanjeh, Nonreciprocal quantum batteries, Phys. Rev, Lett. 132~(21) (2024) 210402.

\bibitem{Downing2021}
C.~A. Downing, V.~A. Saroka, Exceptional points in oligomer chains, Commun. Phys. 4~(1) (2021) 254.

\bibitem{Downing2022}
C.~A. Downing, T.~J. Sturges, Directionality between driven-dissipative resonators, Europhys. Lett. 140~(3) (2022) 35001.

\bibitem{Downing2023}
C.~A. Downing, A.~Vidiella-Barranco, Parametrically driving a quantum oscillator into exceptionality, Sci. Rep. 13~(1) (2023) 11004.

\bibitem{BenArosh2021}
L.~Ben~Arosh, M.~C. Cross, R.~Lifshitz, Quantum limit cycles and the rayleigh and van der pol oscillators, Phys. Rev. Res. 3 (2021) 013130.

\bibitem{Amitai2018.PhysRevE.97.052203}
E.~Amitai, M.~Koppenh\"ofer, N.~L\"orch, C.~Bruder, Quantum effects in amplitude death of coupled anharmonic self-oscillators, Phys. Rev. E 97 (2018) 052203.

\bibitem{Minganti2019}
F.~Minganti, A.~Miranowicz, R.~W. Chhajlany, F.~Nori, Quantum exceptional points of non-hermitian hamiltonians and liouvillians: The effects of quantum jumps, Phys. Rev. A 100~(6) (2019) 062131.

\bibitem{Chimczak2023}
G.~Chimczak, A.~Kowalewska-Kudłaszyk, E.~Lange, K.~Bartkiewicz, J.~Peřina, The effect of thermal photons on exceptional points in coupled resonators, Sci. Rep. 13~(1) (2023) 5859.

\bibitem{Farina2019}
D.~Farina, G.~M. Andolina, A.~Mari, M.~Polini, V.~Giovannetti, Charger-mediated energy transfer for quantum batteries: An open-system approach, Phys. Rev. B 99~(3) (2019) 035421.

\bibitem{Zhang2021}
G.-Q. Zhang, Z.~Chen, W.~Xiong, C.-H. Lam, J.~Q. You, Parity-symmetry-breaking quantum phase transition via parametric drive in a cavity magnonic system, Phys. Rev. B 104~(6) (2021) 064423.

\bibitem{meurer_sympy_2017}
A.~Meurer, C.~P. Smith, M.~Paprocki, O.~{\v{C}}ertík, S.~B. Kirpichev, M.~Rocklin, A.~Kumar, S.~Ivanov, J.~K. Moore, S.~Singh, T.~Rathnayake, S.~Vig, B.~E. Granger, R.~P. Muller, F.~Bonazzi, H.~Gupta, S.~Vats, F.~Johansson, F.~Pedregosa, M.~J. Curry, A.~R. Terrel, {\v{S}}.~Rou\v{c}ka, A.~Saboo, I.~Fernando, S.~Kulal, R.~Cimrman, A.~Scopatz, {SymPy}: symbolic computing in {Python}, PeerJ Comput. Sci. 3 (2017) e103.

\bibitem{blasiak_combinatorics_2005}
P.~Blasiak, Combinatorics of boson normal ordering and some applications (2005).
\newblock \href {http://arxiv.org/abs/quant-ph/0507206} {\path{arXiv:quant-ph/0507206}}.

\bibitem{Mendez_2005}
M.~A. Méndez, P.~Blasiak, K.~A. Penson, Combinatorial approach to generalized bell and stirling numbers and boson normal ordering problem, J. Math. Phys. 46~(8) (2005) 083511.

\bibitem{schlosshauer_decoherence_2007}
M.~Schlosshauer, Decoherence and the quantum-to-classical transition, Springer, 2007.

\bibitem{breuer_theory_2002}
H.-P. Breuer, F.~Petruccione, The theory of open quantum systems, Oxford University Press, 2002.

\bibitem{pikovsky_synchronization_2001}
A.~Pikovsky, M.~Rosenblum, J.~Kurths, Synchronization: a universal concept in nonlinear sciences, Cambridge University Press, 2001.

\bibitem{Bender1998}
C.~M. Bender, S.~Boettcher, Real spectra in non-hermitian hamiltonians having $\mathcal{P}\mathcal{T}$ symmetry, Phys. Rev. Lett. 80~(24) (1998) 5243.

\bibitem{Wiersig2020}
J.~Wiersig, Prospects and fundamental limits in exceptional point-based sensing, Nat. Commun. 11~(1) (2020) 2454.

\bibitem{Zhou2024}
W.~Zhou, J.~Liu, J.~Zhu, D.~Gromyko, C.~Qiu, L.~Wu, Exceptional points unveiling quantum limit of fluorescence rates in non-hermitian plexcitonic single-photon sources, APL Quantum 1~(1) (2024) 016110.

\bibitem{Caloz2018}
C.~Caloz, A.~Alù, S.~Tretyakov, D.~Sounas, K.~Achouri, Z.-L. Deck-Léger, Electromagnetic nonreciprocity, Phys. Rev. Appl. 10~(4) (2018) 047001.

\bibitem{Dek2012}
L.~Deák, T.~F\"{u}l\"{o}p, Reciprocity in quantum, electromagnetic and other wave scattering, Ann. of Phys. 327~(4) (2012) 1050.

\end{thebibliography}

\end{document}